\documentclass[pdflatex,sn-nature]{sn-jnl}
\usepackage{graphicx}%
\usepackage{multirow}%
\usepackage{amsmath,amssymb,amsfonts}%
\usepackage{amsthm}%
\usepackage{mathrsfs}%
\usepackage[title]{appendix}%
\usepackage{xcolor}%
\usepackage{textcomp}%
\usepackage{manyfoot}%
\usepackage{booktabs}%
\usepackage{algorithm}%
\usepackage{algorithmicx}%
\usepackage{algpseudocode}%
\usepackage{listings}%
\usepackage{subfigure}
\usepackage{longtable}
\usepackage{titlesec} % For section formatting

\setcitestyle{super,open={},close={}}
\let\citep\cite
\let\citet\cite

\titleformat{\section}{\normalfont\Large\bfseries}{}{0pt}{}
\titleformat{\subsection}{\normalfont\large\bfseries}{}{0pt}{}
\titleformat{\subsubsection}{\normalfont\normalsize\bfseries}{}{0pt}{}

\begin{document}

%Manually ensure LaTeX creates the .aux file
\typeout{LABEL: fig:bayesian_pca}
\typeout{LABEL: fig:BSE_obs-mod}
\typeout{LABEL: fig:gaussian_mass}
\typeout{LABEL: sec:Methods}
\typeout{LABEL: eq:GOF}

%TC:ignore
\title[Article Title]{Homogeneous accretion of the Earth in the inner Solar System}

\author*[1]{\fnm{Paolo A.} \sur{Sossi}}\email{paolo.sossi@eaps.ethz.ch}

\author[1]{\fnm{Dan J.} \sur{Bower}}\email{dbower@eaps.ethz.ch}

\affil*[1]{\orgdiv{Institute of Geochemistry and Petrology, Department of Earth and Planetary Sciences}, \orgname{ETH Zürich}, \orgaddress{\street{Clausiusstrasse 25}, \city{Zürich}, \postcode{CH-8092}, \state{ZH}, \country{Switzerland}}}

% Dan ORCID: 0000-0002-0673-4860
% Paolo ORCID: 0000-0002-1462-1882

%%==================================%%
%% Sample for unstructured abstract %%
%%==================================%%

\abstract{Meteorites are classified as either non-carbonaceous- (NC) or carbonaceous (CC), representing bodies that likely formed in the inner- or outer solar system, respectively. Despite its location in the inner solar system, the Earth is thought to contain either minor- ($\sim$6~\%) or substantial amounts ($\sim$40~\%) of outer solar system material. However, because neither interpretation leverages variations among multiple isotopic systems simultaneously, Earth's provenance remains equivocal. Here, we examine variations in 10 nucleosynthetic isotope anomalies among planetary- and meteorite parent bodies to show that the linear extension of an array defined by NC bodies in any two isotopic anomalies always intersects the observed isotopic composition of the bulk silicate Earth to within 1 standard deviation. The Earth therefore formed exclusively from inner solar system material whose composition did not vary over the course of accretion and was, on average, unlike that of any chondrite. Extension of the NC array yields isotopic compositions for Mercury and Venus that are more extreme than for Earth, implying a spatial or temporal gradient during the formation of the terrestrial planets.}

\keywords{accretion, earth, isotope, inner solar system}

%%\pacs[JEL Classification]{D8, H51}

%%\pacs[MSC Classification]{35A01, 65L10, 65L12, 65L20, 65L70}

\maketitle

\noindent {\textbf{Published version:} Sossi, P.A. \& Bower, D.J. Homogeneous accretion of the Earth in the inner Solar System. Nature Astronomy (2026).} \href{https://doi.org/10.1038/s41550-026-02824-7}{https://doi.org/10.1038/s41550-026-02824-7}

\newpage
\section{Main Text}

\label{sec1}

The identification of two, distinct populations of meteorites from their mass-independent isotopic compositions \citep{Warren2011}, the `isotopic dichotomy', has precipitated a revolution in our understanding of the provenance of planetary materials, and, in turn, the spatio-temporal evolution of the early solar system \citep{Kruijer2017age,Schiller2018,Yap2023, Rufenacht2023}. These two populations, named `non-carbonaceous' (NC) and carbonaceous (CC), preserve small differences in the abundances of isotopes produced by different nucleosynthetic processes, often expressed as their parts-per-ten-thousand deviation from a standard:

\begin{equation}
    \varepsilon^iX = \left( \frac{(^iX/^jX)_{reservoir}}{(^iX/^jX)_{standard}} - 1, \right) \times 10000
    \label{eq:epsilon}
\end{equation}

where $i$ and $j$ are the isotopic masses and $X$ the element. Isotopic anomalies in $\varepsilon^{54}$Cr and $\varepsilon^{50}$Ti, upon which the dichotomy was first mooted, are roughly linearly correlated among NC bodies, with CI chondrites, a CC body, falling close to \citep{williams2020chondrules} or on an extension of this correlation \citep{Palme2024}. Because the composition of the Earth, as inferred from that of its mantle, the bulk silicate Earth (BSE), is intermediate between the NC group and CI chondrites, it can be made from a mixture of CI chondrites and any combination of NC bodies in $\varepsilon^{54}$Cr-$\varepsilon^{50}$Ti space. The choice of NC body that represents the mixing end-member influences the apparent CI fraction that comprises the BSE. If enstatite chondrites (ECs) are chosen, then the BSE would be made of $\sim$94~\% EC-like- and $\sim$6~\% CI-like matter \citep{Dauphas2017,Dauphas2024,Burkhardt2021,Nimmo2024} whereas if ureilites are chosen, then the BSE must contain 40~\% CI-like material \citep{Schiller2018,Onyett2023,bizzarro2025cosmochemistry}. Because CIs are thought to have originated in the outer solar system, high CI fractions are cited as evidence of substantial accretion of sunward-drifting pebbles to the Earth \citep{Johansen2021pebble,Onyett2023,bizzarro2025cosmochemistry}, whereas low CI fractions are posited to reflect its classical, oligarchic growth from planetesimals largely within the inner solar system \citep{Burkhardt2021,Nimmo2024}.
 
Subsequently, the discovery of the dichotomy in two isotope ratios of a heavy element, Mo, revealed that the BSE is an $s$-process-enriched end-member among all extant meteorite groups \citep{Budde16}. This property was shown to extend to other heavy element nuclides (Zr, Ru and Nd \citep{burkhardt2016,fischer2017ruthenium,render2022}) meaning the Earth cannot reflect mixtures of the known NC and CC bodies alone. Instead, accretion of a third, `missing' component to the Earth \citep{Burkhardt2021} or preferential evaporation of $s$-process-depleted phases in the envelope of the proto-Earth \citep{Onyett2023} have been proposed as possible solutions. 

The acceptance, by some workers, of a higher apparent CC contribution to the BSE budget of Mo \citep[($\sim40\pm$20\%)][]{Spitzer2020,Budde2023,Nimmo2024} than that recorded in Ti and Cr, is taken as support of an earlier notion\citep{schonbachler2010,rubie2011} that the Earth accreted \textit{heterogeneously} \citep{Dauphas2017}. Proxies that are not nucleosynthetic in origin\citep{schonbachler2010,wangbecker2013,alexander2017origin,varasreus2019}, have been cited in support of a CC contribution to the BSE, particularly in the latter stages of its accretion. More recent models grounded in nucleosynthetic isotope variations indicate that Earth initially accreted NC-rich material and became more CC-like over time, prior to the addition of an NC-rich late veneer\citep{fischer2017ruthenium,Nimmo2024,Dauphas2024}. 
This scenario requires that the Earth's mantle equilibrates imperfectly with its core, such that isotopic anomalies in the BSE of more siderophile elements, namely Mo, reflect later stages of Earth's formation than do lithophile elements, like Ti \citep{Dauphas2017}.
In addition to Mo, $\sim$30~\% of the bulk silicate Earth's Zn budget is interpreted to have been delivered by CC material \citep{savage2022,steller2022nucleosynthetic,martins2023}. 
Differences in Ru isotopic compositions between modern-day- and Archean rocks also indicate a degree of heterogeneous accretion, though this may reflect diversity in the compositions of NC bodies rather than a CC contribution\citep{fischer2020ruthenium,messling2025ru}. 
By contrast, a reassessment of Mo isotopic anomalies in the BSE indicates that they are consistent with an NC origin alone within present analytical uncertainties \citep{yokoyama2019,Bermingham2024}, which, if correct, casts doubt on the necessity for the heterogeneous accretion of CC material.

All of the aforementioned models have been motivated by the perceived necessity to make Earth, at least partly, from some mixture of existing meteorite families or their components. However, these interpretations i) are based on only a subset of the measured isotopic anomalies \citep{Nimmo2024,garai2024}, ii) implicitly assume heterogeneous accretion \citep{Dauphas2024} and/or iii) invoke additional factors, namely envelope processing \citep{Onyett2023} or missing components \citep{Burkhardt2021} to extend or modify the isotopic range in meteorites. Here, we examine the composition of the BSE in relation to other meteorite groups in terms of 10 different isotopic anomalies \textit{via} a probabilistic approach that interrogates the data independent of any accretion model for the Earth. 

\subsection{Results}\label{sec:Results}

\subsubsection{PCA and Bayesian latent factor analysis}

We quantify the isotopic relationship between the BSE and other planetary materials (hereinafter `reservoirs') using the means and standard errors of a range of isotopic anomalies ($\varepsilon^{48}$Ca, $\varepsilon^{50}$Ti, $\varepsilon^{54}$Cr, $\varepsilon^{54}$Fe, $\varepsilon^{64}$Ni, $\varepsilon^{66}$Zn, $\varepsilon^{94}$Mo, $\varepsilon^{95}$Mo, $\varepsilon^{96}$Zr and $\varepsilon^{100}$Ru) for each reservoir, tabulated from the public OriginsLab database \citep{Dauphas2024} (see \nameref{sec:Methods}, \textit{Supplementary Section} \ref{sec:All_data}). The isotopic compositions of $\varepsilon^{94}$Mo and $\varepsilon^{95}$Mo in the BSE were recalculated, taking into account the most recent estimates \cite{Budde2023,Bermingham2024} and yield $\varepsilon^{94}$Mo = 0.00$\pm$0.05 and $\varepsilon^{95}$Mo = 0.03$\pm$0.03 (see \ref{fig:94Mo-95Mo-98Mo}). We use the present-day $\varepsilon^{100}$Ru of the bulk silicate Earth\cite{fischer2017ruthenium,fischer2020ruthenium}. We exclude $\Delta^{17}$O, as variations are thought not to be primarily nucleosynthetic\citep{lyonsyoung2005}. Data for Si isotopes are also neglected, owing to uncertainties in correcting for mass-dependent isotopic fractionation\citep{Onyett2023,Dauphas2024}.  

The isotopic systems are classified based on either their \textit{i)} cosmochemical character or \textit{ii)} nucleosynthetic origin. In the first scheme, the iron-loving, siderophile elements (Fe, Ni, Mo, Ru) are distinguished from the silicate-loving, lithophile elements (Ca, Ti, Cr, Zr, Zn), to identify any differences in the isotopic provenance of material brought to the BSE over time \citep{Dauphas2017}. The second scheme separates iron-peak elements (Ca, Ti, Cr, Fe, Ni and Zn) from heavy elements (Mo, Zr and Ru), whose constituent nuclides have distinct nucleosynthetic heritage \citep{clayton1983}, enabling detection of different stellar sources.

All isotope systems, in addition to each subset, are subjected to dimensionality reduction that combines Principal Component Analysis (PCA) as a prior for Bayesian Latent Factor Analysis (B-LFA; see \nameref{sec:Methods}). 
The analysis requires data for the 10 isotopic systems in each reservoir, limiting the analysis to the ureilites (9/10 ratios), Vesta group (9/10 ratios), L, LL (9/10 ratios), H, EL, EH (9/10 ratios), CI, CR, CV, CO, CM, Mars (9/10 ratios) and the BSE. The few missing isotopic ratios were inserted as synthetic priors based on correlations among meteorite groups (\textit{Supplementary Section} \ref{sec:Data_LFA}). Reservoir mean values are weighted averages of samples whose isotopic compositions fall within uncertainty of one another. The Vesta group is an exception; it is a composite reservoir comprised of samples from Vesta, Angrites, Acapulcoites, Lodranites, Mesosiderites and Main Group Pallasites, based on criteria defined in \textit{Supplementary Section} \ref{sec:Data_LFA}. The dataset used for the analysis is given in \ref{tab:Data_LFA}.

\begin{figure}[htbp]
\centering
\includegraphics[width=1\linewidth]{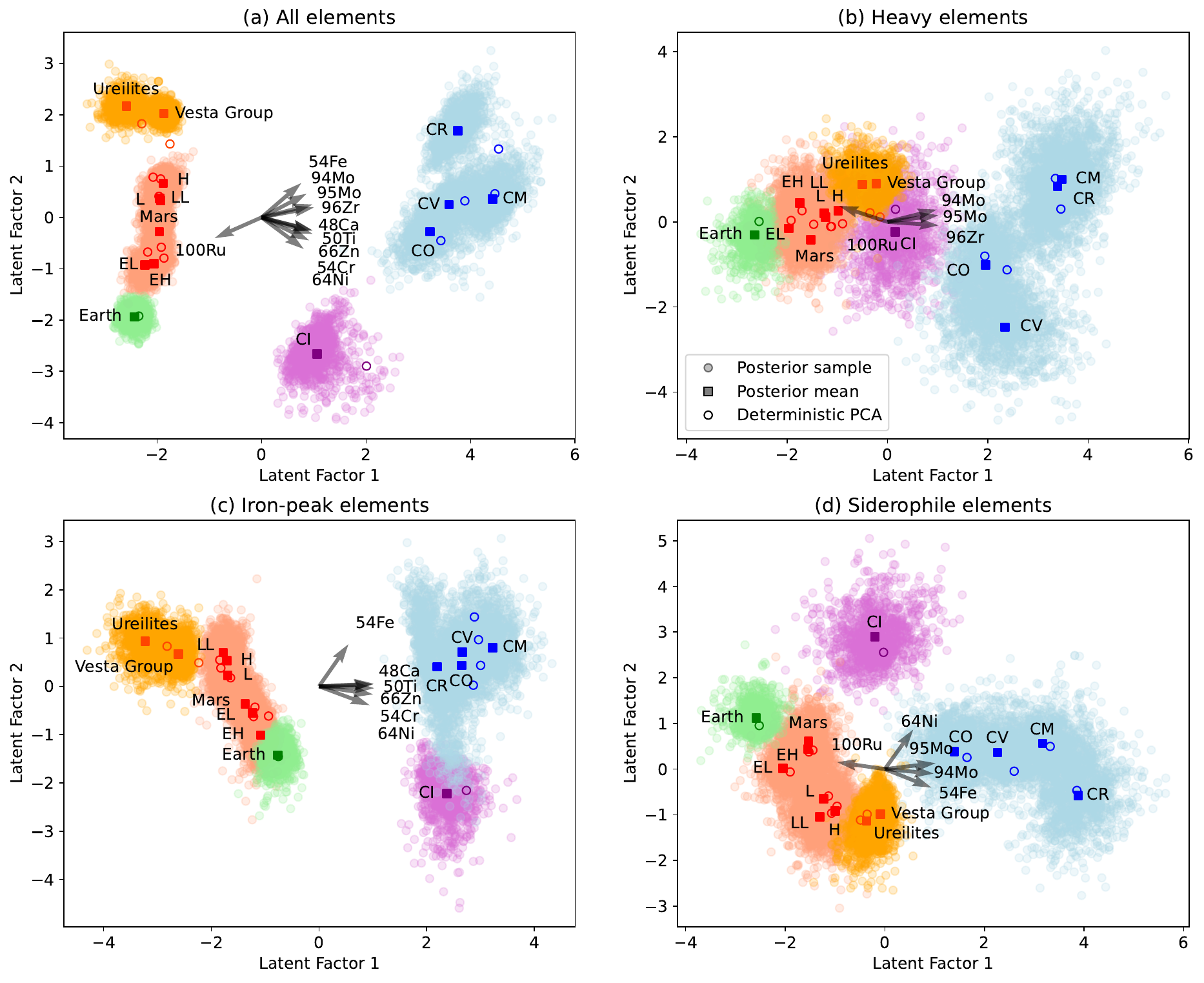}
\caption{\textbf{Results of the Bayesian latent factor analysis (B-LFA) and deterministic Principal Component Analysis (PCA).} Sub-panels show (a) all elements, (b) heavy elements, (c) iron-peak elements, and (d) siderophile elements for all 14 reservoirs. The CI reservoir is clearly separated from the other reservoirs of the CC group. The NC group can also be divided into two subgroups; the ureilities and the Vesta group represent a population distinct from those of the group consisting of H, L, LL (OCs), Mars, EH and EL (ECs), collectively the `OC-EC' subgroup. The bulk silicate Earth is an extension of the `OC-EC' subgroup for all element groups, regardless of their chemical or nucleosynthetic affinity. Open circles are computed from deterministic PCA. Transparent circles show the 10$^4$ samples from the posterior distribution and filled squares are the means of the B-LFA posterior. The arrows denote the loading vector for each isotopic ratio to the latent factors (e.g., a reservoir with high $\varepsilon^{100}$Ru will plot to lower values of LF1 at near-constant LF2 in panel (d)).}
\label{fig:bayesian_pca}
\end{figure}

Based on the means of the isotopic ratios for each reservoir, PCA was performed to determine the latent factors (LFs) that maximise the combined variance along orthogonal axes, ordered by the percentage of the variance they explain (LF1 and LF2; open circles, Fig.~\ref{fig:bayesian_pca}). The PCA is used as the prior distribution for the latent factors and loading matrix of the B-LFA.
In the B-LFA, the likelihood function represents the probability of the observed data (\textit{i.e.,} the 10 isotopic systems in the original, higher-dimensional space) given the latent factors. 
Posterior distributions of latent factors for each reservoir (filled circles, Fig.~\ref{fig:bayesian_pca}) combine the prior PCA estimate with the evidence provided by the observed data and their uncertainties. The B-LFA relaxes the constraint of variance orthogonality imposed by the PCA, and, in so doing, provides more accurate estimates of the means and uncertainties of each reservoir than does PCA alone.

The first two principal components (PC1 and PC2) explain 90 -- 97~\% of the variance in the data, depending on the chosen subset of isotopic anomalies (\ref{tab:PCA_Variance}), substantiating the 2D representation in Fig. \ref{fig:bayesian_pca}. 
Striking is the observation that the NC reservoirs (orange and red shades) are clearly separated from the CC reservoirs (blue and purple shades) by virtue of their distinct LF1 values when all isotopic ratios are considered (Fig. \ref{fig:bayesian_pca}a), a feature that holds for iron-peak elements (Fig. \ref{fig:bayesian_pca}c) and lithophile elements (\ref{fig:bayesian_pca_lithophile}), but is less clearly pronounced among the heavy elements and siderophile elements (Fig. \ref{fig:bayesian_pca}b, d). In all cases, CI chondrites are intermediate to, but distinct from, the NC and CC groups in LF1-LF2 space, providing further support for their status as a singular reservoir \citep{hopp2022ryugu,Yap2023,Dauphas2024,rego2025,shollenberger2025}.

The NC reservoirs occupy a similar location in LF1-LF2 space relative to one another, independent of the subset of isotopic anomalies chosen; the ureilites characterise one extremum of the group, whereas the BSE defines the other. This observation, already evident in $\varepsilon^{54}$Cr-$\varepsilon^{50}$Ti space \citep{Warren2011}, therefore holds across all isotopic systems. Although the relationship between the NC reservoirs has been approximated as linear in $\varepsilon$-$\varepsilon$ space \citep{Burkhardt2021,render2022}, it is evident from Fig. \ref{fig:bayesian_pca} that the Vesta Group and Ureilites (orange points) diverge from other NC reservoirs (see also \ref{fig:SI_LFA-correlations}, ref. \textnormal{\citenum{Rufenacht2023}}). An extension of the line passing through the remaining NC reservoirs; OCs (H, L, LL), Mars and the ECs (EH, EL), hereinafter the `OC-EC' subgroup, \textit{always} intersects the BSE, within uncertainty, for all elements (Fig. \ref{fig:bayesian_pca}a) and any subset thereof (Fig. \ref{fig:bayesian_pca}b-d). By contrast, this same linear extension \textit{never} intersects the isotopic composition of CIs. Therefore, the notion that the BSE can be produced by a mixture of an NC body, namely ECs \citep{Nimmo2024} or ureilites \citep{Schiller2018} and CI chondrites, an interpretation permitted on the basis of $\varepsilon^{54}$Cr-$\varepsilon^{50}$Ti variations alone \citep{Trinquier2009,Palme2024}, is precluded when all isotopic systems are considered together.

The location of any reservoir, $A$, in LF1-LF2 space relative to another reservoir, $B$, can be quantified using the Signed Euclidean Distance, $d^s_{A-B}$, between them, where
\begin{equation}
    d^s_{A-B} = \sqrt{(LF1_A - LF1_B)^2 + (LF2_A - LF2_B)^2} \mathrm{sign}(\vec{a}_B \times \vec{v}_{A-B})
    \label{eq:euclid}
\end{equation}
and the term in brackets is the sign of the cross product of $\vec{a}_B$, the reference axis pointing away from body $B$ and $\vec{v}_{A-B}$, the vector from $B$ to $A$, ensuring $d^s_{A-B}$ can be positive or negative. To render $d^s_{A-B}$ independent of the spread in the subset of isotopic anomalies chosen for the B-LFA, $d^s_{A-B}$ is measured relative to that between a third, fixed reference, $C$ and $B$, $d^s_{C-B}$. The ratio;
\begin{equation}
    R_A = d^s_{A-B}/d^s_{C-B}
    \label{eq:Isotope_euclid}
\end{equation}
defines a dimensionless scale-invariant Signed Euclidean Distance (hereinafter `Isotopic Euclidean Distance'). For $B$ = BSE (hence $R_{BSE} = 0$) and $C$ = OC (hence $R_{OC} = 1$), then for $A$ = EC, $R_{EC}$ yields a constant factor, 0.43$\pm$0.10 across all isotopic anomalies and subsets thereof (Figs. \ref{fig:bayesian_pca}a-d, \ref{tab:Euclidean_Distances}).

Therefore, the BSE, as the silicate portion of a differentiated body, bears a similar isotopic relationship to the undifferentiated, chondritic bodies that constitute the OC-EC subgroup of the NC trend, irrespective of the siderophility or nucleosynthetic origin of the element. Two corollaries emerge from this observation; \textit{i)} Earth formed, on average, from NC material that lies on a linear extension of the trend defined by the `OC-EC' subgroup and \textit{ii)} the composition of this NC material is independent of the isotopic ratio used to infer it. 
Consequently, the isotopic composition of the BSE is intrinsic to the Earth and cannot reflect a mixture of existing planetary materials alone, as indicated for isotopic anomalies in heavy elements\cite{fischer2020ruthenium,Burkhardt2021,messling2025ru} and potentially K isotopes\cite{wang2025}. However, unlike in the aforementioned studies, we propose the end-member status of the BSE among the NC bodies applies equally across each of the 10 isotopic systems considered here.

\subsubsection{Multivariate linear regression}

In order to test hypotheses \textit{i)} and \textit{ii)}, above, we perform linear regressions using the York method \citep{york2004} between the mean of the isotopic anomalies ($x_A$, $y_A$ and their standard errors $\sigma_{x_A}$, $\sigma_{y_A}$) measured in all isotopic systems (e.g., $\varepsilon^{50}$Ti vs. $\varepsilon^{96}$Zr) among reservoirs, $A$, of the `OC-EC' subgroup (see \textit{Supplementary Section} \ref{sec:Data_MLR}, \ref{fig:bayesian_pca_iron_chromium} and \ref{tab:Data_MLR} for the complete list of reservoirs). We fix the measured isotopic anomaly in \textit{one} element, the \textit{predictor}, $x_{BSE}$ in the BSE to compute the isotopic compositions of the remaining nine elements, $y^1_{BSE},~ y^2_{BSE}~...~y^n_{BSE} $ ($n = 9$), from the linear correlations between them and the predictor. Standard errors on the predicted means, $y^n_{BSE}\pm\sigma_{y^n_{BSE}}$, are calculated by Monte Carlo simulation with 10$^4$ samples from normal distributions about the means in the slope ($a\pm\sigma_a$) and intercept ($b\pm\sigma_b$) according to their covariance matrix for each of the linear correlations, and in the measurement of the predictor ($x_{BSE}\pm\sigma_{x_{BSE}}$). This procedure is repeated by cycling through each of the 10 isotopic ratios as the predictor to give the error-weighted mean and standard deviation for each isotopic ratio, $\hat{y}^n_{BSE}\pm \sigma_{\hat{y}^n_{BSE}}$. The analysis is described in the \nameref{sec:Methods} and regression- and Goodness-of-Fit metrics shown in \ref{fig:correlation_matrix} and \ref{fig:gof_matrix}.

\begin{figure}
    \centering
    \includegraphics[width=0.75\linewidth]{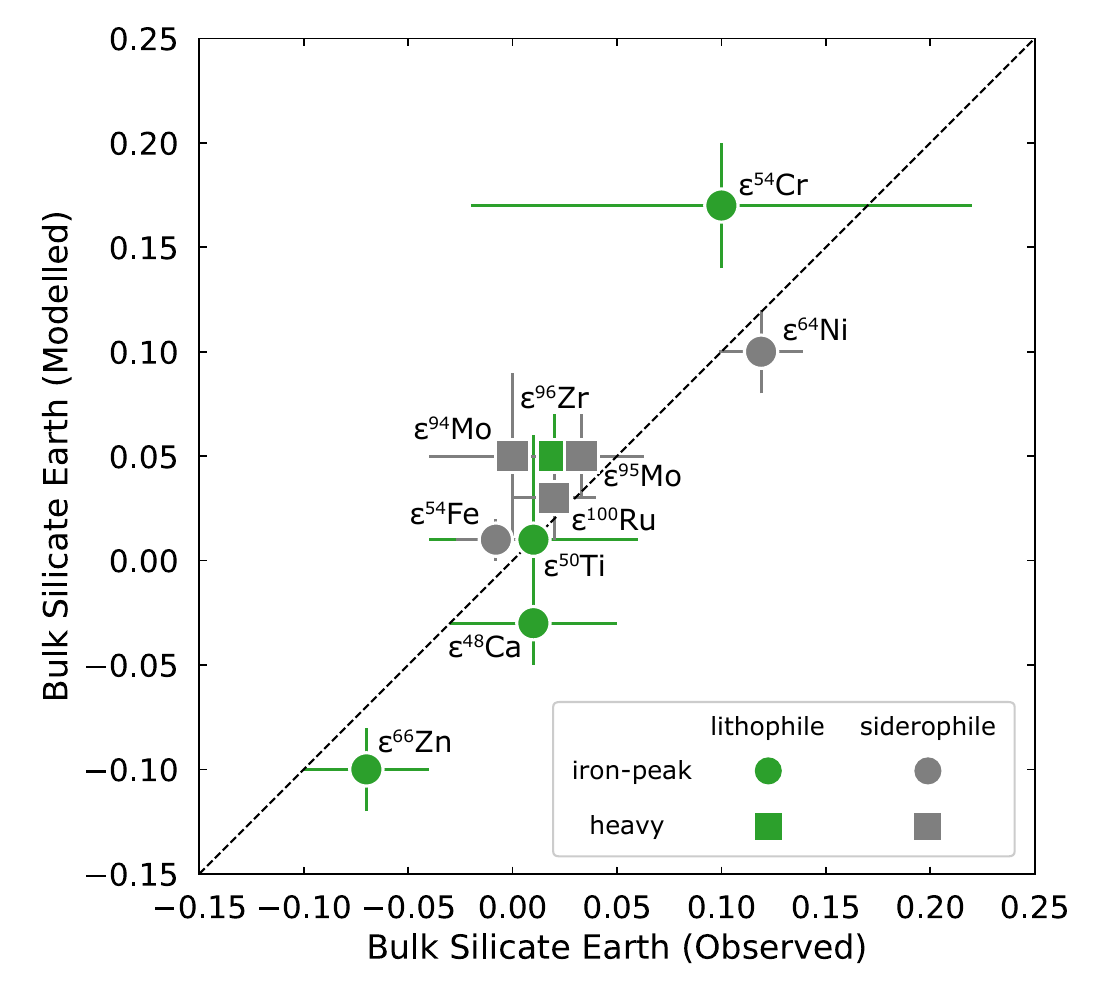}
    \caption{\textbf{Comparison of the observed and modelled isotopic composition of the bulk silicate Earth (BSE).} The observed mean isotopic anomalies in the BSE and their standard errors are shown against those predicted by multivariate linear regression, calculated by assuming the BSE lies on an extension of the NC array defined by the `OC-EC' subgroup, $\hat{y}^n_{BSE}\pm \sigma_{\hat{y}^n_{BSE}}$. Isotopic anomalies are grouped according to the geochemical character and nucleosynthetic origin of the nuclides. The dashed line represents the 1:1 line. All modelled isotopic anomalies fall within 1-$\sigma$ of their observed values.}
    \label{fig:BSE_obs-mod}
\end{figure}

The error-weighted mean of any given predicted isotopic ratio computed across all predictors, $\hat{y}^n_{BSE}\pm \sigma_{\hat{y}^n_{BSE}}$, falls within one standard deviation (mean across all groups is 0.65$\pm$0.55) of the corresponding measured isotopic ratio, $x^n_{BSE}\pm\sigma_{x^n_{BSE}}$ (Fig. \ref{fig:BSE_obs-mod}, \ref{tab:predicted_BSE}, \ref{tab:z-scores}). 
Moreover, the composition of the BSE is predicted equally well using linear regressions between any two combinations of lithophile-, siderophile-, iron-peak- and heavy elements according to their $Z$-scores (\ref{tab:z-scores-collated}, \ref{fig:z-score}). As groups, the mean absolute $Z$-scores per group range between 0.73$\pm$0.18 between heavy element-iron-peak pairs and 0.57$\pm$0.14 between siderophile-siderophile element pairs. This is evident in Fig. \ref{fig:BSE_obs-mod}, where all element groups have isotopic compositions that adhere to the 1:1 line. All individual binary correlations in $\varepsilon$-$\varepsilon$ space are shown in \ref{fig:48Ca_pred} to \ref{fig:100Ru_pred}. Fig. \ref{fig:BSE_obs-mod} shows that the composition of the BSE, for each isotopic system, is entirely consistent with that of an NC body; specifically, an end-member of the `OC-EC' subgroup.

Two alternative hypotheses state that CI chondrites constitute $\sim$6~\% \citep{Nimmo2024, Dauphas2024} or $\sim$40~\% \citep{Onyett2023} by mass of the Earth, with the remainder being made up of an NC body, either ECs \citep{Burkhardt2021,Nimmo2024} or ureilites \citep{Schiller2018,Onyett2023}, respectively. We show that both models are unnecessary for two reasons. First, binary mixing of CI chondrites and an NC body alone cannot reproduce the isotopic composition of heavy elements ($\varepsilon^{94}$Mo, $\varepsilon^{95}$Mo, $\varepsilon^{96}$Zr and $\varepsilon^{100}$Ru) in the BSE, given it is an end-member among all known meteorites for these systems\cite{Budde16,render2022}. Second, while this shortfall can be remedied by a `missing NC-component'\cite{Burkhardt2021,Dauphas2024} or envelope processing \cite{Onyett2023} to produce an end-member rich in $s$-process isotopes, we show that this component, due to the correlated nature of isotopic anomalies in the `OC-EC' subgroup of the NC group, must simultaneously be enriched in neutron-rich isotopes of iron-peak elements, obviating the need for a CI component.

To constrain the maximum permissible mass fraction of CI material, $f$CI, that remains consistent, within uncertainty, with the measured BSE composition, we perturb it by adding CI chondrites. The isotopic compositions for each element in the resulting mixtures are calculated by i) weighting the CI and BSE mass fractions by their corresponding concentrations, and ii) assuming identical concentrations (see \nameref{sec:Methods}).
The BSE's depletion of siderophile elements makes their isotopic compositions, especially those of Mo and Ru, sensitive to CI chondrite addition, where 
$f$CI $<$0.3$\pm$0.1~\% or $<$0.1$^{+0.05}_{-0.03}$~\%, respectively (\ref{fig:SI_f_CI_bymass}). For mixing between isotopically CI- and BSE-like end-members, both with equivalent concentrations of each element, $\varepsilon^{48}$Ca constrains $f$CI to below 2$\pm$0.2~\% in the BSE, and, as a refractory lithophile element, in the bulk Earth (\ref{fig:SI_f_CI_byelement}). In either case, these are lower than previously inferred\cite{Burkhardt2021,Onyett2023,Nimmo2024}. 
Because the isotopic compositions of $\varepsilon^{100}$Ru and $\varepsilon^{48}$Ca in other CC meteorites are more distinct from the BSE than are CI, the allowed $f$CC is correspondingly lower. Since these are \textit{maximum} values, the isotopic composition of the BSE is consistent with lower amounts of CC material. Indeed, $\sim$0.1~\%, $\sim$0.3~\% and 1~\% of CI addition (by mass) would deliver the entire BSE complement of N, C and H, respectively\cite{broadley2022}. Their mantle isotopic compositions, albeit mass-dependent, are also consistent with an NC-like provenance, though their surface reservoirs are fractionated \citep{piani2020}. Thus, the mass of carbonaceous material likely constitutes less than $\sim$0.1~\% of the BSE and $<$2~\% of the bulk Earth.

Excesses in the $\varepsilon^{100}$Ru of Eoarchean  (0.22$\pm$0.04)\cite{fischer2020ruthenium} and Hawaiian rocks (0.09$\pm$0.03)\cite{messling2025ru} relative to the modern-day BSE (0.02$\pm$0.02) implies some heterogeneous accretion to the Earth, supported by $\varepsilon^{40}$K data\cite{wang2025}. These authors suggest such a signature could represent the pre-late veneer mantle, which then accreted an $s$-process-depleted component (either NC or CC)\cite{messling2025ru,wang2025}. However, because
both siderophile \textit{and} lithophile elements predict the modern-day BSE $\varepsilon^{100}$Ru composition equally accurately (Fig. \ref{fig:BSE_obs-mod}), the material responsible for determining the isotopic compositions of elements whose abundances cannot reflect the late veneer (e.g., Ti or Ca), also set the Ru budget of the modern-day BSE. If the interpretation of refs. \citenum{messling2025ru} and \citenum{wang2025} is correct, then the observed consistency between the present-day $\varepsilon^{100}$Ru of the BSE and those of lithophile element isotopic anomalies would be a coincidence.
Alternatively, we propose that $s$-process heterogeneities preserved in $\varepsilon^{100}$Ru\cite{fischer2020ruthenium,messling2025ru} were derived from reservoir(s) whose contribution, on a mass basis, was insignificant relative to that recorded in the contemporary BSE.

\subsection{Discussion}

Our analysis shows that all elements, irrespective of their geochemical character or nucleosynthetic origin, record the same isotopic provenance in the bulk silicate Earth; that of an end-member among the NC group. The composition of the BSE is therefore defined as \textit{homogeneous} with respect to isotopic anomalies. This observation permits two interpretations; \textit{i)} the Earth accreted material that, on average, maintained the same nucleosynthetic isotope composition during the time interval over which its core formed \citep[$\sim$34~Myr][]{kleinewalker2017} (see ref. \citenum{messling2025ru} for an alternative interpretation), and/or \textit{ii)} Earth could have accreted material with distinct nucleosynthetic isotope compositions 
(i.e., heterogeneous accretion), but this heterogeneity must have been subsequently erased through perfect core-mantle equilibration. In both scenarios, the isotopic composition of any given siderophile element recorded in the BSE is equal to that of the bulk Earth.

That the isotopic compositions the `OC-EC' subgroup (including the BSE) in any two elements of differing volatility (e.g., Zn and Ti) define \textit{linear} trends\cite{render2022} indicates that the variations likely arose by mixing between two end-member isotopic reservoirs, $A$ and $B$, in which the concentrations of two elements, $i$ and $j$, were subequal (i.e., $[i/j]_A \sim [i/j]_B$). For a linear relationship to hold, option \textit{ii)} is subject to the additional constraint that the degree of volatile depletion (e.g., Zn/Ti ratio) was subequal among the different nucleosynthetic components that Earth accreted throughout its formation.
While possible\cite{messling2025ru,wang2025}, these additional conditions make option \textit{i)} more probable.

Option \textit{i)} permits that the isotopic composition of Earth-forming material was either \textit{a)} uniform or \textit{b)} adhered to some distribution whose mean corresponds to the composition of the Earth. Since the 'OC-EC subgroup' reservoirs are linearly related in $\varepsilon-\varepsilon$ space, each individual reservoir could have sampled different portions of some continuous, yet imperfectly mixed isotopic distribution. Such isotopic continuity might be achieved across either time (an evolving inner disk composition sampled sequentially by different NC bodies) or space (a spatial gradient in composition sampled contemporaneously\cite{render2017cosmic,fischer2020ruthenium}).

\begin{figure}[!ht]
    \centering
    \includegraphics[width=1\linewidth]{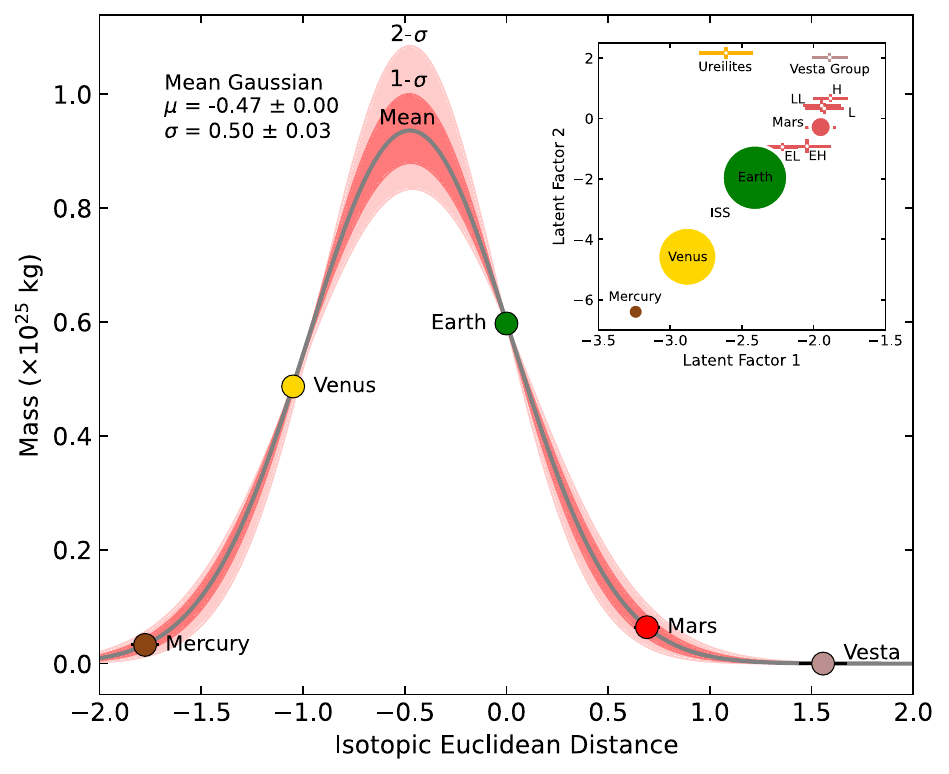}
    \caption{\textbf{Prediction of the isotopic compositions of Venus and Mercury in Isotopic Euclidean Distance space.} The mean and associated 1-$\sigma$ and 2-$\sigma$ confidence intervals around a mass-conserving Gaussian fit to the Isotopic Euclidean distances ($R_A$) of Mars and the Earth as a function of their mass. The confidence intervals were calculated by 10$^4$ Monte Carlo simulations that account for the uncertainties in the isotopic compositions of three bodies across all isotopic anomalies in LF1-LF2 space shown in Fig. 1a. The Isotopic Euclidean Distances of Venus and Mercury are predicted based on their known masses. The peak of the distribution, $\mu = -0.47\pm0.00$ represents the mean isotopic composition of the inner Solar System, expressed in Isotopic Euclidean Distance from the Earth, and $\sigma = 0.50\pm0.03$ its standard deviation and associated standard errors on both values. Inset: The locations of NC bodies in Latent Factor 1 - Latent Factor 2 space across all elements (Fig. 1a), in which the size of the point is scaled proportional to the mass of the body. ISS denotes the mean composition of the inner solar system.}
    \label{fig:gaussian_mass}
\end{figure}

The Isotopic Euclidean Distances ($R_A$, \ref{tab:Euclidean_Distances}) between the Earth (0), Mars (0.69$\pm$0.07) and Vesta (1.55$\pm$0.07), are correlated with their semi-major axes (1.00 au, 1.52 au and 2.36 au, respectively). Although such a gradient has been speculated to exist in individual $\varepsilon$-$\varepsilon$ space, the coordinates of Mercury and Venus along this axis remain undetermined \cite{render2017cosmic,fischer2020ruthenium,mezger2020}. Because the present-day distribution of mass in the inner Solar System varies as a Gaussian function of heliocentric distance about 0.896 au, we propose $R_A$ and planetary mass is also Gaussian. This model enables prediction of the isotopic compositions of Mercury and Venus by requiring the integral of the Gaussian equals the combined masses of the four terrestrial planets. Figure \ref{fig:gaussian_mass} shows the Isotopic Euclidean Distances, $R_{Venus}$ and $R_{Mercury}$, are -1.05$^{+0.01}_{-0.00}$ and -1.78$\pm$0.07, respectively, corresponding to coordinates [LF1, LF2] of [-2.88$\pm$0.00, -4.58$\pm$0.02]$_{Venus}$ and [-3.24$\pm$0.03, -6.40$\pm$0.18]$_{Mercury}$ in Fig. \ref{fig:bayesian_pca} (see also \ref{tab:Venus_Mercury_Isos} and \ref{fig:Venus_Mercury_Isos}).

This scenario predicts that the isotopic compositions of Venus and Mercury lie to more extreme values than Earth. 
One possibility is incomplete mixing took place in the nebular gas, before condensation of grains was able to modify volatile/refractory element ratios. Dynamically, this could relate to turbulent mixing and/or diffusive transport in the early stages of disk accretion due to infall \citep{Kuznetsova_2022}, where temperatures at the midplane in the terrestrial planet-forming region still exceeded the condensation temperatures of oxides and silicates\citep{marschallmorbidelli2023}. 
The Gaussian mass distribution could be achieved either by a Grand Tack-like scenario \citep{walsh2011} or through pressure bumps in the disk that lead to the formation of a ring \citep{izidoro2022planetesimal,Morby22}. Should the terrestrial planets have accreted from individual `rings' at different heliocentric distances, then the predicted compositions of Venus and Mercury need not adhere to the single Gaussian shown in Fig. \ref{fig:gaussian_mass} but could represent individual Gaussians that can move in Euclidean space.  
In either case, because Venus and Earth together represent the bulk of the mass in the inner solar system, they should bear a strong isotopic resemblance to one another. The corollary is that the peripheral bodies (Mercury, Mars, Vesta) should have more extreme compositions, as they stochastically sample poorly mixed tails (Fig. \ref{fig:gaussian_mass}). Sample return missions to the inner planets, Venus and Mercury, are sorely needed to test these ideas.

\clearpage

%TC:ignore
\section{Methods}
\label{sec:Methods}

\subsection{Data aggregation and curation}
Isotopic data were sourced from the compilation maintained by the Origins Lab \citep[hereinafter OLC][]{Dauphas2024}, the sole exception being Mo isotopic data for an estimate of the bulk silicate Earth were also supplemented by two, more recent papers \citep{Budde2023,Bermingham2024} (see \textit{Supplementary Section} \ref{sec:Mo_BSE_Fe}). 
The error-weighted mean, $\mu_i$ and its standard error $\sigma_i$ for each reservoir were calculated from the raw, published data by  defining a weighting factor, $\omega_i$ for each individual sample, $a$, of a given isotope ratio, $i$ (after \cite{york2004}):

\begin{equation}
    \omega^a_i = \frac{1}{({\sigma^a_i})^2}.
    \label{eq:fit_weight}
\end{equation}

The sum across all samples of the products of the weighting factors ($\omega^a_i$) and the measured values ($x^a_i$), divided by the sum of the weighting factors, for each individual sample yields the group mean value, 

\begin{equation}
    \mu_i = \frac{\sum_a \omega^a_{i} x^a_i}{\sum_a \omega^a_{i}}
    \label{eq:predicted_average_SI}
\end{equation}

and the associated standard error:

\begin{equation}
    \sigma_i = \sqrt{\frac{1}{\sum_a \omega^a_{i}}}.
    \label{eq:predicted_sterr_SI}
\end{equation}

In order to detect and remove outliers from the group mean, Grubbs' Test is implemented, in which the weighted standard deviation is calculated;

\begin{equation}
    \sigma_w = \sqrt{\frac{\sum_a \omega^a_{i} (x^a_i - \mu_i)^2}{\sum_a \omega^a_{i}}},
    \label{eq:predicted_stdev_SI}
\end{equation}

and a Grubbs Test statistic is calculated for each sample $x^a_i$;

\begin{equation}
G = \frac{|x^a_i - \mu_i|}{\sigma_w}
    \label{eq:predicted_Grubbs_SI}
\end{equation}

and compared to the critical value, here defined for $\alpha = 0.05$;

\begin{equation}
G_{crit} = \frac{(n-1)t_{\alpha/2n,n-2}}{\sqrt{n(n-2 +t^2_{\alpha/2n,n-2})}}
    \label{eq:predicted_Grubbs_critical_SI}
\end{equation}

where $t_{\alpha/2n,n-2}$ is the critical value from the $t$-distribution with $n-2$ degrees of freedom. If $G > G_{crit}$, the data point is rejected and the test is iterated until there are no outliers remaining. 

\subsection*{Principal Component Analysis / Bayesian-Latent Factor Analysis}

A deterministic Principal Component Analysis (PCA) was performed using the \texttt{PCA} class of the machine-learning library \texttt{sklearn.decomposition} \citep[version 0.24.2][]{scikit-learn} on the means of the isotopic compositions of the reservoirs (\ref{tab:Data_LFA}) to obtain the principal components (PCs) that maximise the combined variance along orthogonal axes (\ref{tab:PCA_Variance}). PCA provides a deterministic solution for dimensionality reduction but does not incorporate measurement uncertainties. To address this limitation, we implemented a Bayesian-Latent Factor Analysis (B-LFA) \citep[][]{Bishop1998,Tipping1999} using \texttt{PyMC5}, which explicitly models uncertainties in the isotopic composition of each reservoir. In this framework, the PCs, which lack associated uncertainty, serve as initial estimates (priors) for the latent factors (scores) and the loading matrix. Unlike PCA, which assumes a fixed solution, B-LFA allows for probabilistic inference by treating the latent factors as a multivariate normal distribution, where the mean is derived from the deterministic PCA solution. To ensure flexibility, each latent factor was assumed to have unit variance (i.e., independent and standardised). Additionally, a non-informative Gamma distribution was used as a hyperprior on the precision (inverse variance) of the loading matrix, enabling the model to adjust uncertainty independently for each latent factor.

To further improve robustness, we modelled the observed data with a Student’s t-distribution, which mitigates sensitivity to outliers. The mean of this distribution was defined as the matrix product of the loading matrix and the latent factors, and the noise scales were set using the standard deviations derived from the isotope data (\ref{tab:Data_LFA}). To accommodate both nearly normal and heavy-tailed data, the normality parameter of the Student’s t-distribution was assigned an exponential prior, \textit{e.g.,} ref. \textnormal{\citenum{Kruschke2013}}. For inference, we performed 12000 steps to sample the posterior distribution. In summary, by leveraging a Bayesian approach, our method extends PCA by incorporating uncertainty, allowing for a more accurate and probabilistic characterisation of isotopic compositions across different reservoirs.

\subsection*{York Multivariate Linear Regression}
We employed the York approach \citep{wehrsaleska2017} to compute linear regressions among isotopic ratios in reservoirs with means $x$ and $y$ and standard errors $\sigma_x$ and $\sigma_y$ following the widely used method of ref. \textnormal{\citenum{york2004}} (their eq. 13):

\begin{subequations}
    \begin{align}
    a = \bar{y} - b\bar{x} \\
    b = \frac{\sum W \beta V}{\sum W \beta U} \\
    \sigma^2_a = \frac{1}{\sum W} + \bar{x}^2\sigma^2_b \\
    \sigma^2_b = \frac{1}{\sum W u^2}.
    \end{align}
    \label{eq:York_fit}
\end{subequations}

where $a$ and $b$ are the intercept and slope of the line and $\sigma_a$ and $\sigma_b$ their associated standard errors and $u = x - x_{adj}$ where $x_{adj}$ is the adjusted value of $x$ according to the fit. The term $\beta$;

\begin{equation}
    \beta = W \left( \frac{U}{\omega(y)} + \frac{bV}{\omega(x)} - (bU + V)\frac{r}{\alpha} \right),
    \label{eq:fit_beta}
\end{equation}

adjusts the contributions of individual data points based on the relative uncertainties in $x$ and $y$ ($\omega$, eq. \ref{eq:fit_weight}), where the regression weighting factor, $W$ is

\begin{equation}
W = \frac{\omega(x)\omega(y)}{\omega(x) + b^2\omega(y)-2br\alpha},
    \label{eq:fit_W}
\end{equation}

and $U = x - \bar{x}$, $V = y - \bar{y}$ (where $\bar{x}$ and $\bar{y}$ denote the predicted values), $r$ is the correlation coefficient between errors in $x$ and $y$, and $\alpha$ is the geometric mean in the standard errors of $x$ and $y$. Here, we set $r = 0$, that is, we impose the condition that errors in the means of two isotopic ratios, $x$ and $y$, are uncorrelated. This simplifies eqs. \ref{eq:fit_beta} and \ref{eq:fit_W}, such that the best-fit values of $a$ and $b$ are symmetrical.

The goodness-of-fit is quantified by the metric `GOF', defined as:

\begin{equation}
    GOF = \frac{\sum_i W_i (y_i - a - bx_i)^2}{n-2} = \frac{\chi^2}{\nu},
    \label{eq:GOF}
\end{equation}

where the term in brackets yields the residuals (the difference between the observed and predicted values of $x$ and $y$) and $n$ the number of data points used in the regression.

\subsubsection{Prediction of the composition of the bulk silicate Earth}

Here, we postulate that the composition of the bulk silicate Earth (BSE) is consistent with a linear extension of the array defined by reservoirs in the `OC-EC' subgroup of NC meteorites in $\varepsilon-\varepsilon$ ($x-y$) space. To test this hypothesis, we fix the isotopic composition of the BSE in \textit{one} of the 10 isotopic systems considered here, termed the `predictor', $x_{pred}$, and compute the remaining nine isotopic ratios ($n = 9$) according to:

\begin{equation}
    \bold{y} = \bold{b} \cdot (x_{pred}\pm \sigma_{x_{pred}}) + \bold{a},
    \label{eq:matrix_multivariate}
\end{equation}

where $\bold{y}$, $\bold{b}$ and $\bold{a}$ are the column vectors containing the resulting $y$ values and the input $b$ (slope) and $a$ (intercept) values as well as their associated standard errors ($\sigma$);

\begin{equation}
\mathbf{y} =
\begin{bmatrix}
y_1 \pm \sigma_{y_1}\\
y_2 \pm \sigma_{y_2}\\
\vdots \\
y_n \pm \sigma_{y_n}
\end{bmatrix},
\quad
\mathbf{b} =
\begin{bmatrix}
b_1 \pm \sigma_{b_1}\\
b_2 \pm \sigma_{b_2}\\
\vdots \\
b_n \pm \sigma_{b_n}
\end{bmatrix},
\quad
\mathbf{a} =
\begin{bmatrix}
a_1 \pm \sigma_{a_1} \\
a_2 \pm \sigma_{a_2} \\
\vdots \\
a_n \pm \sigma_{a_n}
\end{bmatrix}.
\end{equation}

In order to calculate the uncertainties associated with the unknown $\bold{y}$, a Monte Carlo simulation is performed in which the values of $x_{pred}$, $\bold{b}$ and $\bold{a}$ are all varied over 10$^4$ iterations according to an assumed normally distributed $\sigma$ about their mean values. 
In order to leverage \textit{all} regressions to compute the value of the BSE, the weighted mean of the predicted $y$ value of isotope ratio $j$, $\hat{y}_j$, is calculated based on the $y_j$ values given by the other nine predictor isotope ratios ($x_{i,pred}, i \neq j$), following eqs. \ref{eq:fit_weight} to \ref{eq:predicted_sterr_SI}. 

\subsection{Permissible fraction of CI material in the BSE}

This exercise is designed to determine the maximum allowed fraction of CC material that can be added to the bulk silicate Earth, assuming it represents an NC body. To do so, the mass balance equation is solved in two ways. The first uses concentration-weighted mass fractions of the present-day BSE and CI chondrites \cite{palmeoneill2014} to compute the isotopic composition of the mixture:

\begin{equation}
\varepsilon_{\text{BSE+CI},i} = w_{\text{BSE},i} \cdot (\varepsilon_{\text{BSE},i} \pm \sigma_{\text{BSE},i}) +  w_{\text{CI},i} \cdot (\varepsilon_{\text{CI},i} \pm \sigma_{\text{CI},i})
\label{eq:CI_mass_balance_conc}
\end{equation}

where $\varepsilon_{\text{BSE},i}$ and $\varepsilon_{\text{CI},i}$ is the isotopic composition of element $i$ in the bulk silicate Earth or CI chondrites, respectively, and

\begin{equation}
w_{\text{BSE}, i} = \left(\frac{c_{\text{BSE}, i} \cdot f_{\text{BSE}, i}}{c_{\text{BSE}, i} \cdot f_{\text{BSE}, i} + c_{\text{CI}, i} \cdot f_{\text{CI}, i} }\right), w_{\text{CI}, i} = \left(\frac{c_{\text{CI}, i} \cdot f_{\text{CI}, i}}{c_{\text{BSE}, i} \cdot f_{\text{BSE}, i} + c_{\text{CI}, i} \cdot f_{\text{CI}, i} }\right)
\label{eq:CI_mass_balance_weight}
\end{equation}

are the concentration-weighted mass fractions of the BSE and CI chondrites, respectively, in the mixture. Here, $c$ denotes the concentration of $i$ and $f_{\text{BSE}, i} = 1 - f_{\text{CI}, i}$ denotes the relative mass fraction of BSE in the mixture contributing to $i$. The second method is designed such that the concentration of element, $i$, is identical in the BSE-like and in CI-like end-members, that is, $c_{\text{BSE},i} = c_{\text{CI},i}$ for all $i$, for which eq. \ref{eq:CI_mass_balance_conc} simplifies to:

\begin{equation}
\varepsilon_{\text{BSE+CI}, i} = f_{\text{BSE}, i} \cdot (\varepsilon_{\text{BSE}, i} \pm \sigma_{\text{BSE}, i}) + f_{\text{CI}, i} \cdot (\varepsilon_{\text{CI}, i} \pm \sigma_{\text{CI}, i})
\label{eq:CI_mass_balance}
\end{equation}

 The value of $f_{\text{CI}}$ is sampled from 0 to 1 in 10000 steps, and, at each $f_{\text{CI}}$ we performed $N = 10000$ Monte Carlo simulations, where the isotopic compositions $\varepsilon_{\text{BSE}}$ and $\varepsilon_{\text{CI}}$ for the 10 isotopic ratios (i.e., $\varepsilon^{48}$Ca, $\varepsilon^{50}$Ti, $\varepsilon^{54}$Cr, $\varepsilon^{54}$Fe, $\varepsilon^{64}$Ni, $\varepsilon^{66}$Zn, $\varepsilon^{96}$Zr, $\varepsilon^{94}$Mo, $\varepsilon^{95}$Mo, $\varepsilon^{100}$Ru) were randomly drawn from normal distributions centered on their measured values with standard deviations corresponding to their measurement standard errors ($\sigma_{\text{BSE}}$ and $\sigma_{\text{CI}}$). 

To determine the acceptable CI mass fraction for each $N$, the computed mixture composition for each element $i$ must satisfy:

\begin{equation}
\varepsilon_{\text{BSE}, i} - \sigma_{\text{BSE}, i} \leq \varepsilon_{\text{BSE+CI}, i} \leq \varepsilon_{\text{BSE}, i} + \sigma_{\text{BSE}, i}
\end{equation}

where $\varepsilon_{\text{BSE+CI}, i}$ is the simulated mixture composition for element $i$. A simulation was considered successful only if this condition held for all elements simultaneously. The fraction of successful models as a function of $f_{\text{CI}}$ yields an upper bound on the permissible CI fraction in the BSE.

\subsection{Prediction of the isotopic compositions of Venus and Mercury}

Leveraging the observations that i) the masses of the terrestrial planets are distributed in an approximately Gaussian manner about 0.9 au as a function of semi-major axis, and ii) the semi-major axis is correlated with the Isotopic Euclidean Distance ($R_A$) of a body, $A$ ($A$ = Earth (BSE), Mars and Vesta) computed according to eqs. \ref{eq:euclid} and \ref{eq:Isotope_euclid}, we predict the Isotopic Euclidean Distances of Venus ($R_{Venus}$) and Mercury ($R_{Mercury}$). To do so, we fit a mass-conserving Gaussian to the masses of Earth and Mars given their determined values of $R^s_A$ across all isotopic systems (\ref{tab:Euclidean_Distances});

\begin{equation}
    M(R_A) = \sum M_{bodies} \cdot \frac{2}{\sigma \sqrt{2\pi}} \exp\left( -\frac{(R_A - \mu)^2}{2\sigma^2} \right) 
    \label{eq:mass-conserving-gaussian_fit}
\end{equation}

where $\sum M_{bodies}$ is the combined masses of Mercury, Venus, Earth and Mars, $\mu$ is the mean of the Gaussian distribution and $\sigma$ its standard deviation. This procedure is repeated 10$^4$ times by Monte Carlo sampling over normal distributions about the mean of the values of $R_A$ of the Earth and Mars in order to assess their influence on best-fit values of $\mu$ and $\sigma$. The values of $R_{Venus}$ and  $R_{Mercury}$ are recovered by inserting their known values of $M$ into eq. \ref{eq:mass-conserving-gaussian_fit} and solving for $R_A$.  

\backmatter

\section*{Declarations}

\bmhead{Data availability} 
All data generated in this work are available in the main text, in the supplementary information, or as online annexes.  

\bmhead{Code availability}
The combined Principal Component - Bayesian Latent Factor Analysis was performed using the open-source Python package Bedroc v0.2.0, which is available at https://github.com/ExPlanetology/bedroc. The multivariate linear regression analysis and Gaussian fitting model are available as Python scripts at the associated OSF repository at https://doi.org/10.17605/OSF.IO/DH9AK.

\bmhead{Supplementary information}

Supplementary Information can be found online.

\bmhead{Acknowledgements}

We thank Greg Brennecka, Mario Fischer-Gödde and three anonymous reviewers for constructive comments on this work that made for a more rigorous and robust analytical approach and a clearer presentation of the results. PAS thanks Maria Schönbächler, Frédéric Moynier, Raphael Marschall and Alessandro Morbidelli for discussions.
This work was supported by Swiss State Secretariat for Education, Research and Innovation (SERI) under contract No. MB22.00033, a SERI-funded ERC Starting Grant ``2ATMO” (PAS, DJB) and the Swiss National Science Foundation (SNSF) through an Eccellenza Professorship \#203668 (PAS).

\bmhead{Author contribution}

PAS conceived the study, ran the PCA, developed the multivariate linear regression analysis and Gaussian fitting model and wrote the paper. DJB developed the PCA and B-LFA and contributed to writing the paper. 

\bmhead{Competing interests}

The authors declare no competing interests.

\clearpage

%TC:endignore


\noindent


\clearpage

\begin{thebibliography}{10}
\expandafter\ifx\csname url\endcsname\relax
  \def\url#1{\burl{#1}}\fi
\expandafter\ifx\csname urlprefix\endcsname\relax\def\urlprefix{URL }\fi
\providecommand{\bibinfo}[2]{#2}
\providecommand{\eprint}[2][]{\url{#2}}
\providecommand{\doi}[1]{\url{https://doi.org/#1}}
\bibcommenthead

\bibitem{Warren2011}
\bibinfo{author}{Warren, P.~H.}
\newblock \bibinfo{title}{Stable-isotopic anomalies and the accretionary assemblage of the {E}arth and {M}ars: A subordinate role for carbonaceous chondrites}.
\newblock \emph{\bibinfo{journal}{Earth Planet. Sci. Lett.}} \textbf{\bibinfo{volume}{311}}, \bibinfo{pages}{93--100} (\bibinfo{year}{2011}).

\bibitem{Kruijer2017age}
\bibinfo{author}{Kruijer, T.~S.}, \bibinfo{author}{Burkhardt, C.}, \bibinfo{author}{Budde, G.} \& \bibinfo{author}{Kleine, T.}
\newblock \bibinfo{title}{Age of {J}upiter inferred from the distinct genetics and formation times of meteorites}.
\newblock \emph{\bibinfo{journal}{Proc. Natl. Acad. Sci. {U.S.A.}}} \textbf{\bibinfo{volume}{114}}, \bibinfo{pages}{6712--6716} (\bibinfo{year}{2017}).

\bibitem{Schiller2018}
\bibinfo{author}{Schiller, M.}, \bibinfo{author}{Bizzarro, M.} \& \bibinfo{author}{Fernandes, V.~A.}
\newblock \bibinfo{title}{Isotopic evolution of the protoplanetary disk and the building blocks of {E}arth and the {M}oon}.
\newblock \emph{\bibinfo{journal}{Nature}} \textbf{\bibinfo{volume}{555}}, \bibinfo{pages}{507--510} (\bibinfo{year}{2018}).

\bibitem{Yap2023}
\bibinfo{author}{Yap, T.~E.} \& \bibinfo{author}{Tissot, F. L.~H.}
\newblock \bibinfo{title}{The {NC}-{CC} dichotomy explained by significant addition of {CAI}-like dust to the {B}ulk {M}olecular {C}loud ({BMC}) composition}.
\newblock \emph{\bibinfo{journal}{Icarus}} \textbf{\bibinfo{volume}{405}}, \bibinfo{pages}{115680} (\bibinfo{year}{2023}).

\bibitem{Rufenacht2023}
\bibinfo{author}{R\"{u}fenacht, M.} \emph{et~al.}
\newblock \bibinfo{title}{Genetic relationships of solar system bodies based on their nucleosynthetic {Ti} isotope compositions and sub-structures of the solar protoplanetary disk}.
\newblock \emph{\bibinfo{journal}{Geochim. Cosmochim. Acta}} \textbf{\bibinfo{volume}{355}}, \bibinfo{pages}{110--125} (\bibinfo{year}{2023}).

\bibitem{williams2020chondrules}
\bibinfo{author}{Williams, C.~D.} \emph{et~al.}
\newblock \bibinfo{title}{Chondrules reveal large-scale outward transport of inner {S}olar {S}ystem materials in the protoplanetary disk}.
\newblock \emph{\bibinfo{journal}{Proc. Natl. Acad. Sci. {U.S.A.}}} \textbf{\bibinfo{volume}{117}}, \bibinfo{pages}{23426--23435} (\bibinfo{year}{2020}).

\bibitem{Palme2024}
\bibinfo{author}{Palme, H.} \& \bibinfo{author}{Mezger, K.}
\newblock \bibinfo{title}{Nucleosynthetic isotope variations in chondritic meteorites and their relationship to bulk chemistry}.
\newblock \emph{\bibinfo{journal}{Meteorit. Planet. Sci.}} \textbf{\bibinfo{volume}{59}}, \bibinfo{pages}{382--394} (\bibinfo{year}{2024}).

\bibitem{Dauphas2017}
\bibinfo{author}{Dauphas, N.}
\newblock \bibinfo{title}{The isotopic nature of the {E}arth’s accreting material through time}.
\newblock \emph{\bibinfo{journal}{Nature}} \textbf{\bibinfo{volume}{541}}, \bibinfo{pages}{521--524} (\bibinfo{year}{2017}).

\bibitem{Dauphas2024}
\bibinfo{author}{Dauphas, N.}, \bibinfo{author}{Hopp, T.} \& \bibinfo{author}{Nesvorn\'{y}, D.}
\newblock \bibinfo{title}{Bayesian inference on the isotopic building blocks of {M}ars and {E}arth}.
\newblock \emph{\bibinfo{journal}{Icarus}} \textbf{\bibinfo{volume}{408}}, \bibinfo{pages}{115805} (\bibinfo{year}{2024}).

\bibitem{Burkhardt2021}
\bibinfo{author}{Burkhardt, C.} \emph{et~al.}
\newblock \bibinfo{title}{Terrestrial planet formation from lost inner solar system material}.
\newblock \emph{\bibinfo{journal}{Sci. Adv.}} \textbf{\bibinfo{volume}{7}}, \bibinfo{pages}{eabj7601} (\bibinfo{year}{2021}).

\bibitem{Nimmo2024}
\bibinfo{author}{Nimmo, F.}, \bibinfo{author}{Kleine, T.}, \bibinfo{author}{Morbidelli, A.} \& \bibinfo{author}{Nesvorny, D.}
\newblock \bibinfo{title}{Mechanisms and timing of carbonaceous chondrite delivery to the {E}arth}.
\newblock \emph{\bibinfo{journal}{Earth Planet. Sci. Lett.}} \textbf{\bibinfo{volume}{648}}, \bibinfo{pages}{119112} (\bibinfo{year}{2024}).

\bibitem{Onyett2023}
\bibinfo{author}{Onyett, I.~J.} \emph{et~al.}
\newblock \bibinfo{title}{Silicon isotope constraints on terrestrial planet accretion}.
\newblock \emph{\bibinfo{journal}{Nature}} \textbf{\bibinfo{volume}{619}}, \bibinfo{pages}{539--544} (\bibinfo{year}{2023}).

\bibitem{bizzarro2025cosmochemistry}
\bibinfo{author}{Bizzarro, M.}, \bibinfo{author}{Johansen, A.} \& \bibinfo{author}{Dorn, C.}
\newblock \bibinfo{title}{The cosmochemistry of planetary systems}.
\newblock \emph{\bibinfo{journal}{Nat. Rev. Chem.}} \textbf{\bibinfo{volume}{9}}, \bibinfo{pages}{1--19} (\bibinfo{year}{2025}).

\bibitem{Johansen2021pebble}
\bibinfo{author}{Johansen, A.} \emph{et~al.}
\newblock \bibinfo{title}{A pebble accretion model for the formation of the terrestrial planets in the solar system}.
\newblock \emph{\bibinfo{journal}{Sci. Adv.}} \textbf{\bibinfo{volume}{7}}, \bibinfo{pages}{eabc0444} (\bibinfo{year}{2021}).

\bibitem{Budde16}
\bibinfo{author}{Budde, G.} \emph{et~al.}
\newblock \bibinfo{title}{Molybdenum isotopic evidence for the origin of chondrules and a distinct genetic heritage of carbonaceous and non-carbonaceous meteorites}.
\newblock \emph{\bibinfo{journal}{Earth Planet. Sci. Lett.}} \textbf{\bibinfo{volume}{454}}, \bibinfo{pages}{293--303} (\bibinfo{year}{2016}).

\bibitem{burkhardt2016}
\bibinfo{author}{Burkhardt, C.} \emph{et~al.}
\newblock \bibinfo{title}{A nucleosynthetic origin for the {E}arth’s anomalous $^{142}\mathrm{Nd}$ composition}.
\newblock \emph{\bibinfo{journal}{Nature}} \textbf{\bibinfo{volume}{537}}, \bibinfo{pages}{394--398} (\bibinfo{year}{2016}).

\bibitem{fischer2017ruthenium}
\bibinfo{author}{Fischer-G{\"o}dde, M.} \& \bibinfo{author}{Kleine, T.}
\newblock \bibinfo{title}{Ruthenium isotopic evidence for an inner {S}olar {S}ystem origin of the late veneer}.
\newblock \emph{\bibinfo{journal}{Nature}} \textbf{\bibinfo{volume}{541}}, \bibinfo{pages}{525--527} (\bibinfo{year}{2017}).

\bibitem{render2022}
\bibinfo{author}{Render, J.}, \bibinfo{author}{Brennecka, G.~A.}, \bibinfo{author}{Burkhardt, C.} \& \bibinfo{author}{Kleine, T.}
\newblock \bibinfo{title}{Solar {S}ystem evolution and terrestrial planet accretion determined by {Zr} isotopic signatures of meteorites}.
\newblock \emph{\bibinfo{journal}{Earth Planet. Sci. Lett.}} \textbf{\bibinfo{volume}{595}}, \bibinfo{pages}{117748} (\bibinfo{year}{2022}).

\bibitem{Spitzer2020}
\bibinfo{author}{Spitzer, F.} \emph{et~al.}
\newblock \bibinfo{title}{Isotopic evolution of the inner solar system inferred from molybdenum isotopes in meteorites}.
\newblock \emph{\bibinfo{journal}{Astrophys. J. Lett.}} \textbf{\bibinfo{volume}{898}}, \bibinfo{pages}{L2} (\bibinfo{year}{2020}).

\bibitem{Budde2023}
\bibinfo{author}{Budde, G.}, \bibinfo{author}{Tissot, F.~L.}, \bibinfo{author}{Kleine, T.} \& \bibinfo{author}{Marquez, R.~T.}
\newblock \bibinfo{title}{Spurious molybdenum isotope anomalies resulting from non-exponential mass fractionation}.
\newblock \emph{\bibinfo{journal}{Geochem.}} \textbf{\bibinfo{volume}{83}}, \bibinfo{pages}{126007} (\bibinfo{year}{2023}).

\bibitem{schonbachler2010}
\bibinfo{author}{Sch{\"o}nb{\"a}chler, M.}, \bibinfo{author}{Carlson, R.}, \bibinfo{author}{Horan, M.}, \bibinfo{author}{Mock, T.} \& \bibinfo{author}{Hauri, E.}
\newblock \bibinfo{title}{Heterogeneous accretion and the moderately volatile element budget of {E}arth}.
\newblock \emph{\bibinfo{journal}{Science}} \textbf{\bibinfo{volume}{328}}, \bibinfo{pages}{884--887} (\bibinfo{year}{2010}).

\bibitem{rubie2011}
\bibinfo{author}{Rubie, D.~C.} \emph{et~al.}
\newblock \bibinfo{title}{Heterogeneous accretion, composition and core--mantle differentiation of the earth}.
\newblock \emph{\bibinfo{journal}{Earth Planet. Sci. Lett.}} \textbf{\bibinfo{volume}{301}}, \bibinfo{pages}{31--42} (\bibinfo{year}{2011}).

\bibitem{wangbecker2013}
\bibinfo{author}{Wang, Z.} \& \bibinfo{author}{Becker, H.}
\newblock \bibinfo{title}{Ratios of {S}, {S}e and {T}e in the silicate {E}arth require a volatile-rich late veneer}.
\newblock \emph{\bibinfo{journal}{Nature}} \textbf{\bibinfo{volume}{499}}, \bibinfo{pages}{328--331} (\bibinfo{year}{2013}).

\bibitem{alexander2017origin}
\bibinfo{author}{Alexander, C. M.~O.}
\newblock \bibinfo{title}{The origin of inner solar system water}.
\newblock \emph{\bibinfo{journal}{Philos. Trans. R. Soc. A Math. Phys. Eng. Sci.}} \textbf{\bibinfo{volume}{375}}, \bibinfo{pages}{20150384} (\bibinfo{year}{2017}).

\bibitem{varasreus2019}
\bibinfo{author}{Varas-Reus, M.~I.}, \bibinfo{author}{K{\"o}nig, S.}, \bibinfo{author}{Yierpan, A.}, \bibinfo{author}{Lorand, J.-P.} \& \bibinfo{author}{Schoenberg, R.}
\newblock \bibinfo{title}{Selenium isotopes as tracers of a late volatile contribution to {E}arth from the outer {S}olar {S}ystem}.
\newblock \emph{\bibinfo{journal}{Nature Geosci.}} \textbf{\bibinfo{volume}{12}}, \bibinfo{pages}{779--782} (\bibinfo{year}{2019}).

\bibitem{savage2022}
\bibinfo{author}{Savage, P.~S.}, \bibinfo{author}{Moynier, F.} \& \bibinfo{author}{Boyet, M.}
\newblock \bibinfo{title}{Zinc isotope anomalies in primitive meteorites identify the outer solar system as an important source of {E}arth's volatile inventory}.
\newblock \emph{\bibinfo{journal}{Icarus}} \textbf{\bibinfo{volume}{386}}, \bibinfo{pages}{115172} (\bibinfo{year}{2022}).

\bibitem{steller2022nucleosynthetic}
\bibinfo{author}{Steller, T.}, \bibinfo{author}{Burkhardt, C.}, \bibinfo{author}{Yang, C.} \& \bibinfo{author}{Kleine, T.}
\newblock \bibinfo{title}{Nucleosynthetic zinc isotope anomalies reveal a dual origin of terrestrial volatiles}.
\newblock \emph{\bibinfo{journal}{Icarus}} \textbf{\bibinfo{volume}{386}}, \bibinfo{pages}{115171} (\bibinfo{year}{2022}).

\bibitem{martins2023}
\bibinfo{author}{Martins, R.}, \bibinfo{author}{Kuthning, S.}, \bibinfo{author}{Coles, B.~J.}, \bibinfo{author}{Kreissig, K.} \& \bibinfo{author}{Rehk{\"a}mper, M.}
\newblock \bibinfo{title}{Nucleosynthetic isotope anomalies of zinc in meteorites constrain the origin of earth’s volatiles}.
\newblock \emph{\bibinfo{journal}{Science}} \textbf{\bibinfo{volume}{379}}, \bibinfo{pages}{369--372} (\bibinfo{year}{2023}).

\bibitem{fischer2020ruthenium}
\bibinfo{author}{Fischer-G{\"o}dde, M.} \emph{et~al.}
\newblock \bibinfo{title}{Ruthenium isotope vestige of {E}arth’s pre-late-veneer mantle preserved in {A}rchaean rocks}.
\newblock \emph{\bibinfo{journal}{Nature}} \textbf{\bibinfo{volume}{579}}, \bibinfo{pages}{240--244} (\bibinfo{year}{2020}).

\bibitem{messling2025ru}
\bibinfo{author}{Messling, N.} \emph{et~al.}
\newblock \bibinfo{title}{Ru and {W} isotope systematics in ocean island basalts reveals core leakage}.
\newblock \emph{\bibinfo{journal}{Nature}} \textbf{\bibinfo{volume}{642}}, \bibinfo{pages}{376--380} (\bibinfo{year}{2025}).

\bibitem{yokoyama2019}
\bibinfo{author}{Yokoyama, T.}, \bibinfo{author}{Nagai, Y.}, \bibinfo{author}{Fukai, R.} \& \bibinfo{author}{Hirata, T.}
\newblock \bibinfo{title}{Origin and evolution of distinct molybdenum isotopic variabilities within carbonaceous and non-carbonaceous reservoirs}.
\newblock \emph{\bibinfo{journal}{Astrophys. J.}} \textbf{\bibinfo{volume}{883}}, \bibinfo{pages}{62} (\bibinfo{year}{2019}).

\bibitem{Bermingham2024}
\bibinfo{author}{Bermingham, K.~R.} \emph{et~al.}
\newblock \bibinfo{title}{The non-carbonaceous nature of {E}arth’s late-stage accretion}.
\newblock \emph{\bibinfo{journal}{Geochim. Cosmochim. Acta}} \bibinfo{pages}{38--51} (\bibinfo{year}{2025}).

\bibitem{garai2024}
\bibinfo{author}{Garai, S.}, \bibinfo{author}{Olson, P.~L.} \& \bibinfo{author}{Sharp, Z.~D.}
\newblock \bibinfo{title}{Building {E}arth with pebbles made of chondritic components}.
\newblock \emph{\bibinfo{journal}{Geochim. Cosmochim. Acta}} \textbf{\bibinfo{volume}{390}}, \bibinfo{pages}{86--104} (\bibinfo{year}{2024}).

\bibitem{lyonsyoung2005}
\bibinfo{author}{Lyons, J.} \& \bibinfo{author}{Young, E.}
\newblock \bibinfo{title}{{CO} self-shielding as the origin of oxygen isotope anomalies in the early solar nebula}.
\newblock \emph{\bibinfo{journal}{Nature}} \textbf{\bibinfo{volume}{435}}, \bibinfo{pages}{317--320} (\bibinfo{year}{2005}).

\bibitem{clayton1983}
\bibinfo{author}{Clayton, D.~D.}
\newblock \emph{\bibinfo{title}{Principles of {S}tellar {E}volution and {N}ucleosynthesis}}  (\bibinfo{publisher}{University of Chicago Press}, \bibinfo{year}{1983}).

\bibitem{hopp2022ryugu}
\bibinfo{author}{Hopp, T.} \emph{et~al.}
\newblock \bibinfo{title}{Ryugu’s nucleosynthetic heritage from the outskirts of the {S}olar {S}ystem}.
\newblock \emph{\bibinfo{journal}{Sci. Adv.}} \textbf{\bibinfo{volume}{8}}, \bibinfo{pages}{eadd8141} (\bibinfo{year}{2022}).

\bibitem{rego2025}
\bibinfo{author}{Siciliano~Rego, E.}, \bibinfo{author}{Dauphas, N.} \& \bibinfo{author}{Hopp, T.}
\newblock \bibinfo{title}{Consolidating the isotopic trichotomy of planetary materials with new evidence}.
\newblock \emph{\bibinfo{journal}{Geochem. Perspect. Lett.}} \textbf{\bibinfo{volume}{37}}, \bibinfo{pages}{7--11} (\bibinfo{year}{2025}).

\bibitem{shollenberger2025}
\bibinfo{author}{Shollenberger, Q.~R.} \emph{et~al.}
\newblock \bibinfo{title}{Elemental and isotopic signatures of {A}steroid {R}yugu support three early {S}olar {S}ystem reservoirs}.
\newblock \emph{\bibinfo{journal}{Earth Planet. Sci. Lett.}} \textbf{\bibinfo{volume}{664}}, \bibinfo{pages}{119443} (\bibinfo{year}{2025}).

\bibitem{Trinquier2009}
\bibinfo{author}{Trinquier, A.} \emph{et~al.}
\newblock \bibinfo{title}{Origin of nucleosynthetic isotope heterogeneity in the solar protoplanetary disk}.
\newblock \emph{\bibinfo{journal}{Science}} \textbf{\bibinfo{volume}{324}}, \bibinfo{pages}{374--376} (\bibinfo{year}{2009}).

\bibitem{wang2025}
\bibinfo{author}{Wang, D.} \emph{et~al.}
\newblock \bibinfo{title}{Potassium-40 isotopic evidence for an extant pre-giant-impact component of {E}arth’s mantle}.
\newblock \emph{\bibinfo{journal}{Nat. Geosci.}} \bibinfo{pages}{1--6} (\bibinfo{year}{2025}).

\bibitem{york2004}
\bibinfo{author}{York, D.}, \bibinfo{author}{Evensen, N.~M.}, \bibinfo{author}{Mart{\i}nez, M.~L.} \& \bibinfo{author}{De~Basabe~Delgado, J.}
\newblock \bibinfo{title}{Unified equations for the slope, intercept, and standard errors of the best straight line}.
\newblock \emph{\bibinfo{journal}{Am. J. Phys.}} \textbf{\bibinfo{volume}{72}}, \bibinfo{pages}{367--375} (\bibinfo{year}{2004}).

\bibitem{broadley2022}
\bibinfo{author}{Broadley, M.~W.}, \bibinfo{author}{Bekaert, D.~V.}, \bibinfo{author}{Piani, L.}, \bibinfo{author}{F{\"u}ri, E.} \& \bibinfo{author}{Marty, B.}
\newblock \bibinfo{title}{Origin of life-forming volatile elements in the inner {S}olar {S}ystem}.
\newblock \emph{\bibinfo{journal}{Nature}} \textbf{\bibinfo{volume}{611}}, \bibinfo{pages}{245--255} (\bibinfo{year}{2022}).

\bibitem{piani2020}
\bibinfo{author}{Piani, L.} \emph{et~al.}
\newblock \bibinfo{title}{Earth’s water may have been inherited from material similar to enstatite chondrite meteorites}.
\newblock \emph{\bibinfo{journal}{Science}} \textbf{\bibinfo{volume}{369}}, \bibinfo{pages}{1110--1113} (\bibinfo{year}{2020}).

\bibitem{kleinewalker2017}
\bibinfo{author}{Kleine, T.} \& \bibinfo{author}{Walker, R.~J.}
\newblock \bibinfo{title}{Tungsten isotopes in planets}.
\newblock \emph{\bibinfo{journal}{Annu. Rev. Earth Planet. Sci.}} \textbf{\bibinfo{volume}{45}}, \bibinfo{pages}{389--417} (\bibinfo{year}{2017}).

\bibitem{render2017cosmic}
\bibinfo{author}{Render, J.}, \bibinfo{author}{Fischer-G{\"o}dde, M.}, \bibinfo{author}{Burkhardt, C.} \& \bibinfo{author}{Kleine, T.}
\newblock \bibinfo{title}{The cosmic molybdenum-neodymium isotope correlation and the building material of the {E}arth}.
\newblock \emph{\bibinfo{journal}{Geochem. Perspect. Lett}} \textbf{\bibinfo{volume}{3}}, \bibinfo{pages}{170--178} (\bibinfo{year}{2017}).

\bibitem{mezger2020}
\bibinfo{author}{Mezger, K.}, \bibinfo{author}{Sch{\"o}nb{\"a}chler, M.} \& \bibinfo{author}{Bouvier, A.}
\newblock \bibinfo{title}{Accretion of the {E}arth—missing components?}
\newblock \emph{\bibinfo{journal}{Space Sci. Rev.}} \textbf{\bibinfo{volume}{216}}, \bibinfo{pages}{27} (\bibinfo{year}{2020}).

\bibitem{Kuznetsova_2022}
\bibinfo{author}{Kuznetsova, A.}, \bibinfo{author}{Bae, J.}, \bibinfo{author}{Hartmann, L.} \& \bibinfo{author}{Low, M.-M.~M.}
\newblock \bibinfo{title}{Anisotropic infall and substructure formation in embedded disks}.
\newblock \emph{\bibinfo{journal}{Astrophys. J.}} \textbf{\bibinfo{volume}{928}}, \bibinfo{pages}{92} (\bibinfo{year}{2022}).

\bibitem{marschallmorbidelli2023}
\bibinfo{author}{Marschall, R.} \& \bibinfo{author}{Morbidelli, A.}
\newblock \bibinfo{title}{An inflationary disk phase to explain extended protoplanetary dust disks}.
\newblock \emph{\bibinfo{journal}{Astron. Astrophys.}} \textbf{\bibinfo{volume}{677}}, \bibinfo{pages}{A136} (\bibinfo{year}{2023}).

\bibitem{walsh2011}
\bibinfo{author}{Walsh, K.~J.}, \bibinfo{author}{Morbidelli, A.}, \bibinfo{author}{Raymond, S.~N.}, \bibinfo{author}{O'Brien, D.~P.} \& \bibinfo{author}{Mandell, A.~M.}
\newblock \bibinfo{title}{A low mass for {M}ars from {J}upiter’s early gas-driven migration}.
\newblock \emph{\bibinfo{journal}{Nature}} \textbf{\bibinfo{volume}{475}}, \bibinfo{pages}{206--209} (\bibinfo{year}{2011}).

\bibitem{izidoro2022planetesimal}
\bibinfo{author}{Izidoro, A.} \emph{et~al.}
\newblock \bibinfo{title}{Planetesimal rings as the cause of the {S}olar {S}ystem’s planetary architecture}.
\newblock \emph{\bibinfo{journal}{Nat. Astron.}} \textbf{\bibinfo{volume}{6}}, \bibinfo{pages}{357--366} (\bibinfo{year}{2022}).

\bibitem{Morby22}
\bibinfo{author}{Morbidelli, A.} \emph{et~al.}
\newblock \bibinfo{title}{Contemporary formation of early {S}olar {S}ystem planetesimals at two distinct radial locations}.
\newblock \emph{\bibinfo{journal}{Nat. Astron.}} \textbf{\bibinfo{volume}{6}}, \bibinfo{pages}{72 – 79} (\bibinfo{year}{2022}).

\bibitem{scikit-learn}
\bibinfo{author}{Pedregosa, F.} \emph{et~al.}
\newblock \bibinfo{title}{Scikit-learn: {M}achine {L}earning in {P}ython}.
\newblock \emph{\bibinfo{journal}{J. Mach. Learn. Res.}} \textbf{\bibinfo{volume}{12}}, \bibinfo{pages}{2825--2830} (\bibinfo{year}{2011}).

\bibitem{Bishop1998}
\bibinfo{author}{Bishop, C.}
\newblock \bibinfo{editor}{Kearns, M.}, \bibinfo{editor}{Solla, S.} \& \bibinfo{editor}{Cohn, D.} (eds) \emph{\bibinfo{title}{Bayesian {PCA}}}.
\newblock (eds \bibinfo{editor}{Kearns, M.}, \bibinfo{editor}{Solla, S.} \& \bibinfo{editor}{Cohn, D.}) \emph{\bibinfo{booktitle}{Advances in Neural Information Processing Systems}}, Vol.~\bibinfo{volume}{11} (\bibinfo{publisher}{MIT Press}, \bibinfo{year}{1998}).

\bibitem{Tipping1999}
\bibinfo{author}{Tipping, M.~E.} \& \bibinfo{author}{Bishop, C.~M.}
\newblock \bibinfo{title}{Probabilistic {P}rincipal {C}omponent {A}nalysis}.
\newblock \emph{\bibinfo{journal}{J. R. Stat. Soc. Ser. B Stat. Methodol.}} \textbf{\bibinfo{volume}{61}}, \bibinfo{pages}{611--622} (\bibinfo{year}{1999}).

\bibitem{Kruschke2013}
\bibinfo{author}{Kruschke, J.~K.}
\newblock \bibinfo{title}{Bayesian {E}stimation {S}upersedes the t-{T}est}.
\newblock \emph{\bibinfo{journal}{J. Exp. Psychol. Gen.}} \textbf{\bibinfo{volume}{142}}, \bibinfo{pages}{573603} (\bibinfo{year}{2013}).

\bibitem{wehrsaleska2017}
\bibinfo{author}{Wehr, R.} \& \bibinfo{author}{Saleska, S.~R.}
\newblock \bibinfo{title}{The long-solved problem of the best-fit straight line: application to isotopic mixing lines}.
\newblock \emph{\bibinfo{journal}{Biogeosci.}} \textbf{\bibinfo{volume}{14}}, \bibinfo{pages}{17--29} (\bibinfo{year}{2017}).

\bibitem{palmeoneill2014}
\bibinfo{author}{Palme, H.} \& \bibinfo{author}{O’Neill, H. S.~C.}
\newblock \bibinfo{title}{ in \textit{Cosmochemical estimates of mantle composition}} \bibinfo{edition}{2} edn, (eds \bibinfo{editor}{Holland, H.~D.} \& \bibinfo{editor}{Turekian, K.~K.}) \emph{\bibinfo{booktitle}{Treatise on {G}eochemistry}}, Vol.~\bibinfo{volume}{2} \bibinfo{pages}{1--39} (\bibinfo{publisher}{Elsevier}, \bibinfo{address}{Oxford}, \bibinfo{year}{2014}).

\end{thebibliography}
\end{document}